\documentclass[%
reprint,
superscriptaddress,
aps,
floatfix,
prl
]{revtex4-2}

\usepackage{amsmath,amssymb}
\usepackage{graphicx}
\usepackage{dcolumn}
\usepackage{bm}
\usepackage{xcolor}
\usepackage{hyperref}
\usepackage[capitalize]{cleveref}

\hypersetup{
    colorlinks=false,  
    pdfborder={0 0 0}  
}

\newcommand{\BenchP}{15}

\newcommand{\DenseTensorMemory}{286\,\mathrm{GiB}}

\newcommand{\DenseValidationNmax}{3}
\newcommand{\DenseValidationError}{6.24\times10^{-16}}

\newcommand{\RhsQConvergenceNmax}{12}
\newcommand{\RhsQDefault}{24}
\newcommand{\RhsQReference}{48}
\newcommand{\RhsQDefaultError}{4.10\times10^{-11}}

\begin{document}

\title{Nonlinear Landau Collisions Without Collision Tensors}

\author{R. Jorge}
 \email{rogerio.jorge@wisc.edu}
\author{B. Herfray}
\author{C. Vega}
\author{V. Zhdankin}
\affiliation{Department of Physics, University of Wisconsin–Madison, Madison, Wisconsin 53706, USA}

\begin{abstract}
Far-from-equilibrium plasmas require nonlinear Coulomb collisions, but direct three-dimensional Hermite discretization of the Landau operator needs an impractical dense tensor. By porting quantum chemistry Coulomb-integral methods, we reduce the six-dimensional integrals to one-center Coulomb moments and separable exponential-sum contractions. This gives a four-order-of-magnitude working-memory reduction and enables nonlinear relaxation tests. Numerical simulations preserve invariants and show that finite-basis linearization changes relaxation and produces a fourfold angular error.
\end{abstract}

\maketitle

%
%

While the Boltzmann operator stemming from Coulomb scattering can be put in a Fokker-Planck form \cite{Chandrasekhar1943,Rosenbluth1957}, its differential-integral character makes its numerical implementation challenging.
This has led to different avenues of research, implementing either a linearized version of the collision operator \cite{Braginskii1965,Catto1977,Ji2006,Sugama2017}, an approximate linear model \cite{Frei2021, Abel2008a, Sugama2009}, or other operators that retain mass, momentum, and energy conservation \cite{Lenard1958,Dougherty1964,Rosen2025}.
%
%
Here, by porting quantum-chemistry Coulomb-integral methods, we obtain a numerically stable nonlinear Landau operator for multispecies plasmas that is both physically accurate and numerically efficient.
%

The use of an accurate collision operator is important for both short and long mean-free-path plasmas.
As shown in Ref.~\cite{Pezzi2016}, mean-free-path arguments to justify the use of simplified collision operators fail for non-Maxwellian plasmas, such as \textit{in situ} solar wind measurements \cite{Maruca2013}.
%
%
This is due to the presence of strong velocity gradients, which locally amplify the role of collisions, smoothing those gradients in timescales that can be much shorter than collisional timescales
\cite{Pezzi2016}.
In addition, intermediate collisionality phenomena, such as the edge of fusion devices \cite{Jorge2017}, also involve plasma turbulence where perturbations can reach order unity, highlighting the need for a proper treatment of the nonlinear plasma collision operator.
%
%
Finally, even for low collisionality regimes with linearized operators, the eigenmodes associated with the solution of such systems can differ substantially between collision operators \cite{Jorge2018}.

Basis-function decompositions have provided closed-form collision coefficients for many moment systems \cite{Ji2009,Hirvijoki2016,Pfefferle2017,Jorge2019a,Hunana2022}.
%
%
Due to the tensorial character of such decompositions \cite{Grad1949} and the nonlinearities appearing in the collision operator, the final set of equations usually requires a long table of coefficients, with large sums involving tensorial products and unstable large-number cancellations, which can be numerically inefficient, and even become prohibitive to calculate when the temperature-dependent coefficients vary in time and space.
This involves spherical harmonics, Hermite-Laguerre, tensorial Hermite, and Burnett polynomials, which are not only used to model collisional systems, but also collisionless phenomena such as magnetic reconnection \cite{Loureiro2013a,Numata2015}, cascades in velocity space \cite{Servidio2017}, non-local transport \cite{Joglekar2018}, and magnetic pumping \cite{Lichko2017}.
One of the main obstacles in the use of the nonlinear Landau operator in such simulations is the fact that for $n$ modes per velocity direction, the distribution function has $n^3$ coefficients, while the bilinear collision map has $n^9$ entries.
Here, we show that this tensor is unnecessary, as the six-dimensional Landau integral reduces to one-center Coulomb/Talmi integrals and separable tensor-product contractions.
Therefore, the full multispecies Cartesian-Hermite Landau coefficient tensor can be calculated using one-center Coulomb moments with a separable coupling using the Boys function \cite{PareschiRussoToscani2000,LiWangWang2020,LiRenWang2021}.


In this Letter, we port the Cartesian Gaussian function methods used to solve Coulomb integrals in quantum chemistry to the solution of the nonlinear collision operator in plasma physics.
As we show below, this allows us to obtain a fast, numerically efficient expression for the coefficients of the Hermite expansion of the plasma distribution function, reducing the full multispecies Cartesian-Hermite coefficient tensor to one-center Coulomb moments whose only inter-dimensional coupling is the scalar factor from the Boys function $F_0$ \cite{Boys1950}.
It also leads to a much simpler derivation of the coefficients when compared with tensorial methods.
The main methods employed here are the Brody-Moshinsky (harmonic oscillator) transformation brackets from nuclear physics \cite{Moshinsky1959} and the McMurchie-Davidson integral scheme \cite{McMurchie1978}.
In essence, this amounts to converting the two-body Coulomb integral in six-dimensional velocity space $(\mathbf v, \mathbf v')$ into a single Coulomb integral in three dimensions via a center of mass transformation $\mathbf u=\mathbf v-\mathbf v'$, followed by the conversion of the $1/u$ singularity into a Boys function \cite{Boys1950} commonly used in molecular orbital integrals \cite{Rys1983} and in Schwinger's proper time formalism in quantum electrodynamics (Eq. 3.24 of Ref. \cite{Schwinger1951}).
%
%
We note that the same oscillator algebra can be rotated to Hermite--Laguerre or spherical bases.
Here, we employ three-dimensional Hermite basis functions used to model weakly collisional plasmas \cite{Delzanno2015,Vencels2016,Servidio2017}, such as sub-ion scale turbulence simulations \cite{Roytershteyn2019}.
%
%
%
We also implement the resulting nonlinear operator numerically using exclusively one-dimensional tables and tensor-product contractions, enabling three-dimensional relaxation tests on a single processor core where dense tensor storage is already impractical.


We start from first principles, using the nonrelativistic Boltzmann equation for the evolution of the plasma distribution function $f^a(\mathbf x, \mathbf v, t)$ for each species $a$, dependent on space $\mathbf x$, velocity $\mathbf v$, and time $t$, which can be written as
\begin{equation}\label{eq:boltzmann}
    \frac{\partial f^a}{\partial t}+\mathbf v \cdot \frac{\partial f^a}{\partial \mathbf x}+\frac{\mathbf F^a}{m^a}\cdot \frac{\partial f^a}{\partial \mathbf v}=\sum_b C^{ab}(f^a,f^b),
\end{equation}
where $m^a$ is the particle's mass, $\mathbf F^a=q^a(\mathbf E + \mathbf v \times \mathbf B)$ is the Lorentz force with $q^a$ the charge, $\mathbf E$ the electric field and $\mathbf B$ the magnetic field.
The collision operator $C^{ab}$ can be written using either a Rosenbluth potential formulation \cite{Rosenbluth1957} or using the symmetric tensor $U_{ij}=\partial^2 u / \partial u_i \partial u_j=(\delta_{ij} u^2-u_i u_j)/u^3$ where $\mathbf u=\mathbf v -\mathbf v'$ and $u=|\mathbf u|$ as
\begin{equation}\label{eq:collisionoperator}
    C^{ab} = \frac{\Gamma^{ab}}{2 m^a}\frac{\partial}{\partial v_i}\int d \mathbf v' U_{ij} \left(f^b\frac{\partial f^a}{\partial v_j}-\frac{m^a}{m^b}f^a\frac{\partial f^b}{\partial v'_j}\right),
\end{equation}
with $f^a=f^a(\mathbf v)$, $f^b=f^b(\mathbf v')$, $\Gamma^{ab}=(q^a q^b)^2 \ln \Lambda ^{ab} / 4 \pi \epsilon_0^2 m^a$, $\ln \Lambda^{ab}$ the Coulomb logarithm, and Einstein notation is used to sum over the indices $i$ and $j$.

We now employ a basis function decomposition of the distribution function of the species $s$, $f^s=\sum_{n,m,p} f^{s}_{nmp}(\mathbf x,t) \psi_{nmp}^s(\mathbf v)$ with the time-evolved basis coefficients, $\psi^s_{nmp}$ the product of basis functions $\psi^s_{nmp}=\psi_n(\overline v_{x}^s)\psi_m(\overline v_{y}^s)\psi_p(\overline v_{z}^s)$  with $\overline {\mathbf v}^s = \mathbf v / v_{th}^s$, $v_{th}^s=\sqrt{2 T^s/m^s}$ the thermal velocity and $T^s$ an arbitrary normalization temperature for the species $s$, and single basis functions $\psi_n(x)=H_n(x)e^{-x^2}/\sqrt{2^n n! \pi}=(-1)^n(d^n/dx^n)e^{-x^2}/\sqrt{2^n n! \pi}$ with $H_n(x)$ a Hermite polynomial, and the dual $\Psi_n(x)=H_n(x)/\sqrt{2^n n!}$ so that $\int_{-\infty}^{\infty} \psi_n(x) \Psi_{n'}(x) dx = \delta_{n,n'}$.
Because the construction uses harmonic-oscillator algebra, the same coefficients can be rotated into Hermite--Laguerre polynomials used in gyrokinetic simulations \cite{Mandell2018,Frei2020} or spherical oscillator basis \cite{Hirshman1976} by standard basis changes.

In order to turn \cref{eq:boltzmann} into an equation for the coefficients $f^a_{nmp} \equiv f^a_k$, we multiply \cref{eq:boltzmann} by the dual basis functions $\Psi^{a}_{nmp} \equiv \Psi^a_k$ and integrate over velocity space $\mathbf v$.
This way, the collision operator moments can then be written as a product of distribution function basis coefficients as 
\begin{equation}\label{eq:momentsCab}
    C_k^{ab}=\int d {\mathbf v} C^{ab} \Psi^{a}_{k}=\sum_{k^a}\sum_{k^b}f^a_{k^a}f^b_{k^b}C^{ab}_{k,k^a,k^b},
\end{equation}
where the subscript $k^a$ is the set of Hermite indices $(n^a, m^a, p^a)$ for species $a$, and analogously for $k^b=(n^b, m^b, p^b)$ and $k=(n, m, p)$.
The Hermite index $m^a$ is not to be confused with the mass of species $a$.
Additionally, we integrate by parts the outer $\partial/\partial v_i$ derivative, and expand each distribution function $f^s$ into basis functions $\psi^a_{n^a m^a p^a}(\mathbf v)=\psi^a_{k^a}$ and $\psi^b_{n^b m^b p^b}(\mathbf v')=\psi^b_{k^b}$, yielding
\begin{align}
    C^{ab}_{k,k^a,k^b}=&-\frac{\nu^{ab}}{2}\int d \overline{\mathbf v}^a d \overline {\mathbf v}'^b \frac{\partial \Psi^{a}_{k}}{\partial \overline v_{i}^a}\overline U_{ij}\nonumber\\
    &\times\left(\psi^b_{k^b}\frac{\partial \psi^a_{k^a}}{\partial \overline v_{j}^a}-\frac{m^a}{m^b}\frac{v_{th}^a}{v_{th}^b}\psi^a_{k^a}\frac{\partial \psi^b_{k^b}}{\partial \overline v_{j}^{'b}}\right),\label{eq:Cabkkakb}
\end{align}
where $\nu^{ab}=\Gamma^{ab}(v_{th}^b)^3/m^a$, $\overline U_{ij}=v_{th}^a U_{ij}(\mathbf u)=U_{ij}(\overline{\mathbf u}^a)$ with $\overline{\mathbf u}^a=\mathbf u/v_{th}^a$.
Both $f^a$ and basis functions are assumed to decay fast enough making boundary terms zero.

We now simplify \cref{eq:Cabkkakb} by applying two standard identities: the fact that derivatives of basis functions can be cast as basis functions of a given order (ladder identities), and that the product of basis functions can be cast as a linear combination of basis functions.
This is true for either the Cartesian, spherical, or associated Laguerre decompositions, but we show here the derivation in the Hermite basis due to the simplicity of its derivative (also called ladder) operator.
A spherical decomposition would introduce Clebsch-Gordan and Gaunt coefficients that can also be computed efficiently \cite{Rasch2012}.
We find that $\partial \Psi^{a}_{k}/\partial \overline v_{i}^a=\sqrt{2 k_i}\Psi^{a}_{k-e_i}$ and $\partial \psi^{a}_{k}/\partial \overline v_{i}^a=-\sqrt{2 (k_i+1)}\psi^{a}_{k+e_i}$ with $e_i=(100),(010)$ or $(001)$ if $i=1,2$ or $3$, respectively, and $k_i$ the index $n,m$ or $p$ if $i=1,2$ or $3$, respectively.
The product of basis functions can also be written as $\Psi^{a}_{k} \psi^a_{k^a}=\sum_{k'}P^{k'}_{k,k^a} \psi^a_{k'}$ where the $k'=(n'm'p')$ sum is truncated and the selection rules (which apply in each Cartesian component) impose $P^{k'}_{k,k^a}=0$ unless $k'=k+k^a$(mod 2) and $|k-k^a|\le k' \le k+k^a$.
The factor $P^{k'}_{k,k^a}=P^{n'}_{n,n_a}P^{m'}_{m,m_a}P^{p'}_{p,p_a}$ can be computed directly using the closed form $P^{n'}_{n,n^a}=r!\binom{n}{r}\binom{n^a}{r}\sqrt{n'!/n!n^a!}$ with $r=(n+n^a-n')/2$ or using a stable recurrence.
%
We can then rewrite \cref{eq:Cabkkakb} as
\begin{align}
    C^{ab}_{k,k^a,k^b}=&\nu^{ab}\sum_{k'}\sum_{i,j=1}^3\sqrt{k_i(k_{j}^a+1)}\int d \overline{\mathbf v}^a d \overline {\mathbf v'}^b \psi^{a}_{k'}\overline U_{ij}\nonumber\\
    &\times\left(\psi^b_{k^b}P^{k'}_{k-e_i,k^a+e_j}-P^{ab,ij}_{k,k',k^a,k^b}\psi^b_{k^b+e_j}\right),\label{eq:Cabkkakb2}
\end{align}
where
\begin{equation}
    P^{ab,ij}_{k,k',k^a,k^b}=\frac{m^a}{m^b}\frac{v_{th}^a}{v_{th}^b}\sqrt{\frac{k_{j}^b+1}{k_{j}^a+1}}P^{k'}_{k-e_i,k^a},
\end{equation}
is the coefficient proportional to $\sqrt{T^a m^a/T^b m^b}$ that, together with $\nu^{ab}$, encompasses the multispecies character of the collision operator.
We note that, while the two Hermite products in \cref{eq:Cabkkakb2} generate two independent finite sums, $\sum_{k'} P^{k'}_{k-e_i,k^a+e_j}$ and $\sum_{k''} P^{k''}_{k-e_i,k^a}$, as both sums occur over the same Hermite basis (up to the usual selection rules), we relabel the dummy index $k''$ as $k'$ for compactness.


Leveraging the fact that $\overline U_{ij}$ is only a function of $\overline {\mathbf u}^a= \overline {\mathbf v}^a - (v_{th}^b/v_{th}^a)\overline {\mathbf v'}^b$, the integral in \cref{eq:Cabkkakb2} in $(\overline {\mathbf v}^a, \overline {\mathbf v'}^b)$ coordinates can be simplified by moving to a thermal-speed-weighted relative frame $(\overline {\mathbf u}^{ab}, \overline {\mathbf U}^{ab})$, akin to the center of mass frame.
We therefore define the angle $\theta$ as $\cos\theta=v_{th}^a/v_{th}^{ab}$ and $\sin\theta=v_{th}^b/v_{th}^{ab}$, with $v_{th}^{ab}=\sqrt{(v_{th}^a)^2+(v_{th}^b)^2}$.
Then, the frame is defined as $\overline {\mathbf u}^{ab}=\cos\theta\overline{\mathbf u}^a={\mathbf u}/v_{th}^{ab}$, $\overline {\mathbf U}^{ab}=\sin\theta\overline {\mathbf v}^a+\cos\theta\overline {\mathbf v'}^b$.
The center of mass integral has unit Jacobian $d \overline {\mathbf v}^a d \overline {\mathbf v'}^b=d \overline {\mathbf u}^{ab} d \overline {\mathbf U}^{ab}$, which can be derived from the non-normalized coordinates $d\mathbf v d\mathbf v'=d\mathbf u d \mathbf U$ where $\mathbf u = \mathbf v - \mathbf v'$ and $\mathbf U=\sin^2\theta \mathbf v+\cos^2\theta \mathbf v'=v_{th}^b\cos\theta \overline{\mathbf U}^{ab}$, or from the inverse transformation $\overline{\mathbf v}^a=\sin\theta\overline{\mathbf U}^{ab}+\cos\theta\overline{\mathbf u}^{ab}$ and $\overline{\mathbf v'}^b=\cos\theta\overline{\mathbf U}^{ab}-\sin\theta\overline{\mathbf u}^{ab}$.
We now show that such a frame separates the six-dimensional integral into one three-dimensional Coulomb integral and three independent one-dimensional integrals (one per Cartesian component), which could be done analytically.
This is due to the identity $(\overline v^a)^2+({\overline v'}^b)^{2} = (\overline U^{ab})^2+(\overline u^{ab})^2$ and the Hermite rotation integral identity \cite{Feldheim1940}
\begin{align}
    \int_{-\infty}^{\infty}&e^{-t^2}H_n(t\cos\theta-r \sin\theta)H_m(t \sin\theta + r \cos\theta)dt\nonumber\\
    &=\sqrt{\pi}(-1)^n \cos^m\theta\sin^n\theta H_{m+n}(r),
\end{align}
which is a special case of the Talmi-Moshinsky (harmonic oscillator) transformation brackets \cite{Moshinsky1959}.
The resulting integral in $\overline{\mathbf U}$ collapses to the center of mass ground state
\begin{equation}\label{eq:Uint}
    \int d \overline{\mathbf U}^{ab}\psi^a_{k'}(\overline{\mathbf v}^a)\psi^b_{k^b}(\overline{\mathbf v'}^b)=a_{k' ,k^b}\psi_{K^{'b}}(\overline{\mathbf u}^{ab}),
\end{equation}
%
%
where, in \cref{eq:Uint}, $a_{k', k^b}=(-1)^{|k^b|}(v_{th}^a)^{|k'|}(v_{th}^b)^{|k^b|}\sqrt{(k'+k^b)!/k'!k^b!}/(v_{th}^{ab})^{|k'|+|k^b|}$, $|k|=n+m+p$, $k!=n!m!p!$ and $K^{'b}=k'+k^b$.


The moments of the collision operator, \cref{eq:Cabkkakb2}, can then be written as
\begin{align}
    &C^{ab}_{k,k^a,k^b}=\nu^{ab}\sum_{k'}\sum_{i,j=1}^3\sqrt{k_i(k_{j}^a+1)}\int d \overline{\mathbf u}^{ab}\overline U_{ij}\times\nonumber\\
    &(a_{k',k^b}P^{k'}_{k-e_i,k^a+e_j}\psi_{K^{'b}}-P^{ab,ij}_{k,k',k^a,k^b}a_{k',k^b+e_j}\psi_{K^{'b}+e_j}).\label{eq:Cabkkakb3}
\end{align}
Note that $\overline U_{ij}=\overline U_{ij}(\overline{\mathbf u}^a)=\overline U_{ij}(\overline{\mathbf u}^{ab}/\cos\theta)=\cos\theta\overline U_{ij}(\overline{\mathbf u}^{ab})$.
The remaining integral
\begin{align}
    &G_{K^{'b}}^{ij}=\int d\overline{\mathbf u}^{ab} \overline U_{ij}({\overline{\mathbf u}^{ab}})\psi_{K^{'b}}(\overline{\mathbf u}^{ab})=\sqrt{(K^{'b}_j+1)}\times\nonumber\\
    &\left(\sqrt{K^{'b}_i+\delta_{ij}}I_{K^{'b}+e_j-e_i}+\sqrt{K^{'b}_i+\delta_{ij}+1}I_{K^{'b}+e_j+e_i}\right),
\end{align}
is performed by integrating by parts with $U_{ij}({\mathbf u})=\partial_{u_j}(u_i/u)$ and $\partial_{u_j} \psi_{K^{'b}}(\mathbf u)=-\sqrt{2(K^{'b}_j+1)}\psi_{K^{'b}+e_j}(\mathbf u)$, and using the ladder identity for Hermite polynomials.
This leads to the scalar Coulomb moment $I_K$ that is only nonzero if each component of $K$ is even, with $K=(2a,2b,2c)$ and $A=a+b+c$, computed as
\begin{equation}\label{eq:IK}
    I_K=\int d \mathbf u \frac{\psi_K(\mathbf u)}{|\mathbf u|}=
    \frac{(-1)^A\sqrt{(2a)!(2b)!(2c)!}}{\sqrt{\pi}(2A+1)2^{A-1} a!b!c!},
\end{equation}
which is evaluated using the identity $1/u=(2/\sqrt{\pi})\int_0^\infty e^{-t^2 u^2}dt$ often used for Talmi integrals and for retrieving the Boys function $F_n(T)=\int_0^1 x^{2n}e^{-T x^2}dx$, with $F_n(0)=1/(2n+1)$ the factor leading to the resulting $1/(2A+1)$ factor in \cref{eq:IK}, similar to the one appearing in one-center Coulomb integrals \cite{Helgaker2000}.
The convention $I_L=0$ is used whenever any component of $L$ is negative or odd.
The final result reads
\begin{align}
    &C^{ab}_{k,k^a,k^b}=\cos\theta\nu^{ab}\sum_{k'}\sum_{i,j=1}^3\sqrt{k_i(k_{j}^a+1)}\times(a_{k',k^b}\times\nonumber\\
    &P^{k'}_{k-e_i,k^a+e_j}G_{K^{'b}}^{ij}-P_{k,k',k^a,k^b}^{abij}a_{k',k^b+e_j}G_{K^{'b}+e_j}^{ij}).\label{eq:Cabkkakbfinal}
\end{align}
%
%
Shifted or anisotropic Gaussian centers require the corresponding Boys-function moments, as in standard Gaussian-integral algorithms.


\begin{figure}[t]
    \centering
    \includegraphics[width=0.49\textwidth]{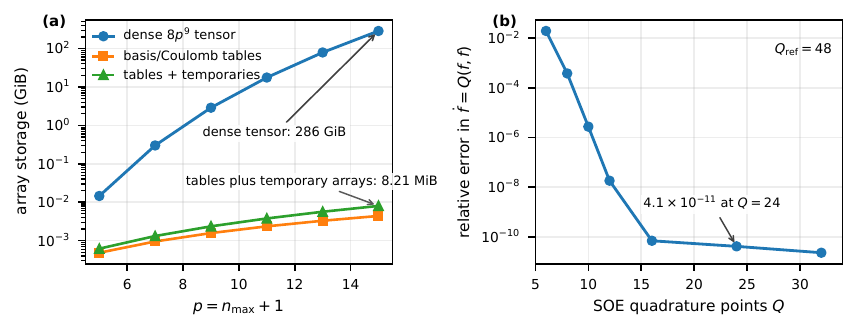}
    \vspace{-0.6cm}
    \caption{
    (a) Storage needs, with dense tensor (blue), and one-center/SOE method with and without temporary variables (green and orange, respectively) in terms of polynomials $n_{\text{max}}$ used.
    (b) Convergence with SOE quadrature points.
    }
    \vspace{-0.6cm}
    \label{fig:tensor_barrier}
\end{figure}

As shown above, our approach reduces the evaluation to a small number of stable contractions and precomputable Coulomb moments $I_K$, as opposed to direct tensor formulas that have many nested sums that scale poorly with the number of moments and may require cancellations between extremely large numbers.
Additionally, the scalar moment $I_K$ is fully separable except for the factor $(2A+1)^{-1}$.
This factor can be separated using a sum of exponentials (SOE)
\begin{equation}
\frac{1}{2A+1}=\int_0^1s^{2A}ds
\approx
\sum_{q=1}^{Q}w_qs_q^{2A}
=
\sum_{q=1}^{Q}w_q
(s_q^{2a})(s_q^{2b})(s_q^{2c}),
\end{equation}
which converts the Coulomb part of the collision map into a sum of Cartesian tensor products.
%
%
In practice, we use \(Q\)-point Gauss--Legendre quadrature on \([0,1]\) where $Q$ is the chosen number of quadrature points to keep.
Due to the fact that \(s_q^{2A}=s_q^{2a}s_q^{2b}s_q^{2c}=e^{(2A)\log s_q}\), the quadrature is a separable exponential-sum approximation \cite{BeylkinMonzon2005}.
The advantage of the method is shown in \cref{fig:tensor_barrier}: direct storage of \(C_{k,k^a,k^b}\) would require \(8n^9\) bytes, \(\DenseTensorMemory\) at \(p=\BenchP\), whereas the measured one-center/SOE tables plus temporary arrays require 8.21 MiB, a factor of $10^4$ reduction.
%
%
%
Where the dense tensor can still be constructed explicitly, \(n_{\max}=\DenseValidationNmax\), contraction with the dense tensor agrees with the tensorized collision update to \(\DenseValidationError\).
The factor \(8\) in \(8n^9\) in \cref{fig:tensor_barrier} is the number of bytes per double-precision coefficient, while the dense tensor has \(n^9=(n^3)^3\) entries.
The SOE quadrature is assessed using maximum Hermite coefficients $n=m=p$, where at \(n_{\max}=\RhsQConvergenceNmax\), \(Q=\RhsQDefault\) agrees with \(Q=\RhsQReference\) to \(\RhsQDefaultError\).
%
%
Dense assembly is already impractical at low orders, as 
at \(p=5\)
it took more than 20 times longer to run than the tensorized collision update at \(p=\BenchP\).
%


The standard conservation laws for the collision operator $C^{ab}$, which should be satisfied by any plasma collision operator, namely energy, momentum and particle number, can be demonstrated for the moments $C^{ab}_{k}=C^{ab}_{nmp}$ by expressing the collision invariants $(1,\mathbf v,v^2)$ in terms of the dual basis (hence, Hermite polynomials) $\Psi^{a}_{k}$, leading to the three coefficient-level identities $C^{ab}_{000}=0$, $m^a v_{th}^a C^{ab}_{e_i}=-m^b v_{th}^b C^{ba}_{e_i}$ and $\sum_{i=1}^3m^a (v_{th}^a)^2C^{ab}_{2e_i}=-\sum_{i=1}^3m^b (v_{th}^b)^2C^{ba}_{2e_i}$.
The first identity follows from the factor \(\sqrt{k_i}\) in \(C^{ab}_{k,k^a,k^b}\). 
The momentum and energy identities follow from the product selection rules and the exchange symmetry \(a\leftrightarrow b\), for which \(\mathbf u\to-\mathbf u\), \(\psi_K(-\mathbf u)=(-1)^{|K|}\psi_K(\mathbf u)\), and \(U_{ij}(-\mathbf u)=U_{ij}(\mathbf u)\).

\begin{figure}
    \centering
    \includegraphics[width=0.99\linewidth]{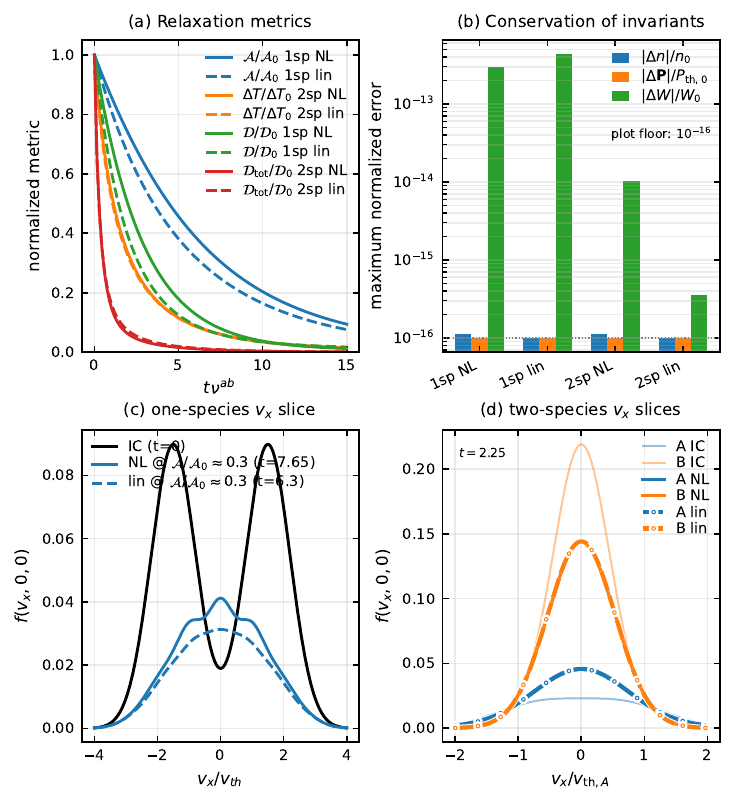}
    \caption{Comparison between linear and nonlinear collisional relaxation using a far-from-equilibrium initial condition. (a) Normalized anisotropy and relative entropy. (b) Maximum relative errors in conserved quantities. (c) One species $v_x$ slice. (d) Two species $v_x$ slice at $t=2.25/\nu^{ab}$. Thirteen Hermite polynomials.}
    \label{fig:panel1}
    \vspace{-0.5cm}
\end{figure}

We now implement the Boltzmann equation, \cref{eq:boltzmann}, in the homogeneous limit (no advection and no external forces), i.e., $\partial_t f^a=\sum_b C^{ab}(f^a,f^b)$, using the newly derived form of the collision operator, \cref{eq:Cabkkakbfinal}.
This allows us to assess collisional relaxation, conservation properties, and the difference between nonlinear and linearized dynamics.
To prevent recurrence and phase-space filamentation, 
%
%
a diagonal high-Hermite filter, proportional to the Hermite modes \(n(n-1)(n-2)\) (and similarly for $m$ and $p$), conserving mass, momentum, and energy \cite{Vencels2016}, is added with a small coefficient $\nu=10^{-4}$.
%
The Coulomb tables are precomputed and the resulting moments are obtained by tensor-product contractions.

\begin{figure}
    \centering
    \includegraphics[width=0.99\linewidth]{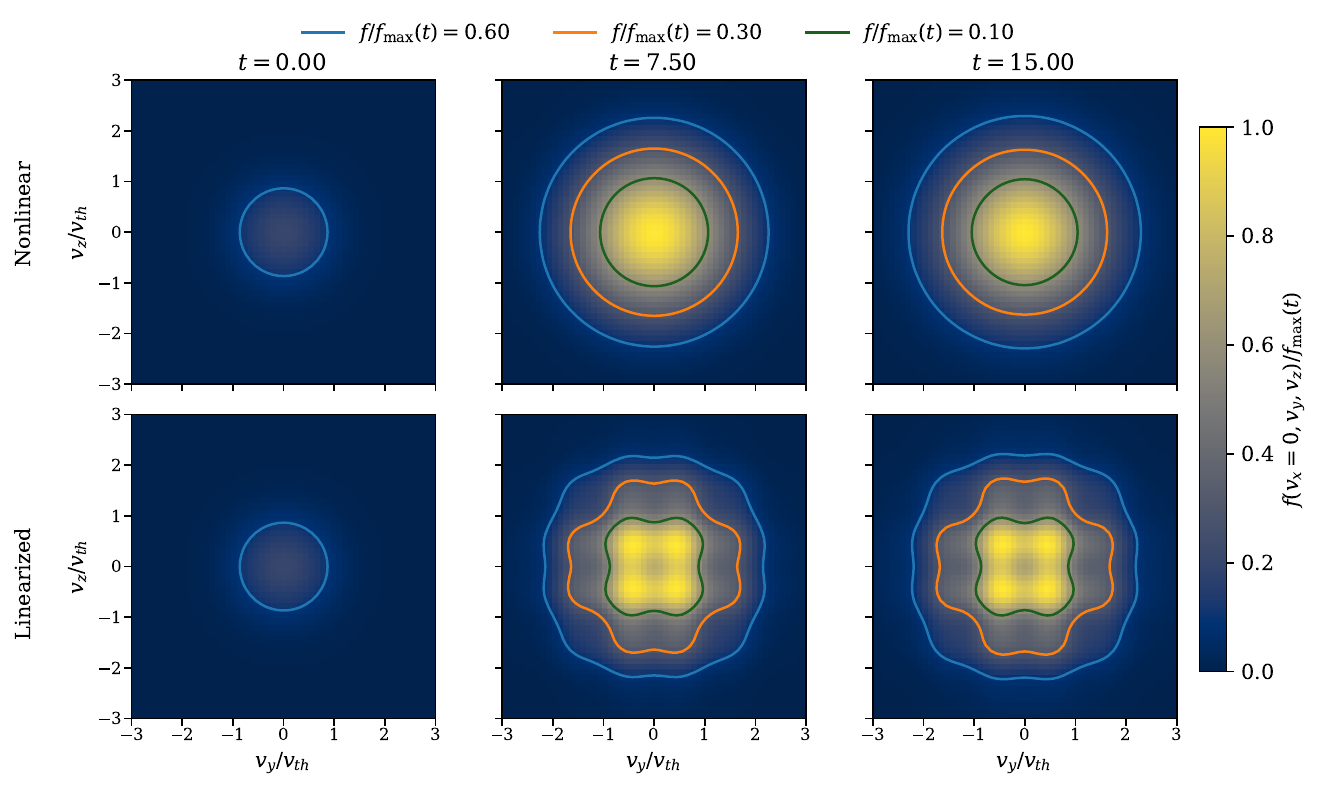}
    \caption{Evolution of the distribution function in the perpendicular $v_y$ and $v_z$ velocity space for the case of the nonlinear (top) and linearized (bottom) collision operators.
    The linearized Cartesian-Hermite truncation produces a finite-basis fourfold distortion.
    }
    \label{fig:panel1_3D}
    \vspace{-0.5cm}
\end{figure}


We show in \cref{fig:panel1} both the linearized (dashed lines), where second-order perturbations away from a Maxwellian $M(v)$ {where $v$ can be any component of $\mathbf v=(v_x,v_y,v_z)$} are neglected, and fully nonlinear (solid lines) results, both for a one-species (bottom left) and a two-species case (bottom right).
%
Panel (B) reports the maximum normalized error on each run, showing good conservation properties.
For the one-species case in \cref{fig:panel1} (a,b), we use a far-from-equilibrium two-stream initial condition,
$f(v_x,v_y,v_z,0)= \tfrac12\!\left[M(v_x-u_0)+M(v_x+u_0)\right]M(v_y)M(v_z)$ with $u_0=1.5$ in units of thermal velocity,
projected onto the truncated Hermite basis with $M(v)=e^{-v^2}/\pi^{1/2}$.
The two-species test uses equal densities, masses $m_B=2 m_A=2$ and temperatures $T_A=1.5 T_B=1.5$.
We then show in units of collisional time $1/\nu^{ab}$ the temperature anisotropy metric $\mathcal{A}=(T_z-T_x)/[(T_x+T_y+T_z)/3]$ and relative entropy $\mathcal D^s=\int f^s \ln(f^s/f^s(t=0)d\mathbf v$ with $\mathcal D_{\text{tot}}=\sum_s \mathcal D^s$.
The temperature for each species $s$ is computed as $T_i^s=2 T^s [\Theta_{2i}^s/\Theta_{0i}^s-(\Theta_{1i}^s/\Theta_{0i}^s)^2]$ with $\Theta_{ni}^s=\int d \overline{\mathbf v} \overline v_i^n f^s$ and $T^s$ the reference temperature used before.
For example, $\Theta_{0x}^s=2[(f_{000}^s/2+f_{200}^s/\sqrt2)/f_{000}^s-(f_{100}^s/(\sqrt2 f_{000}^s))^2]$, and similarly for $y,z$.
As expected, the normalized diagnostics decrease as the distributions relax.
%
%
%
%
The corresponding one-species slice at \(\mathcal{A}/\mathcal{A}_0=0.3\) is shown in \cref{fig:panel1} (c).
%
The conserved energy and momentum invariants are computed as
$P_x=m v_{th}^4 f_{100}/\sqrt{2}$ (similarly for $P_y,P_z$) and
$W={m v_{th}^5}/{2}\!\left[({3}/{2})f_{000}+({f_{200}+f_{020}+f_{002}})/{\sqrt{2}}\right]$,
with $T_a=2W_a/(3n_a)$.
The time evolution is performed using an explicit third-order strong stability preserving Runge-Kutta (SSPRK3) method.
%
%
The invariants are computed from the Hermite moments and remain conserved during the simulation.
For momentum, \(|\Delta\mathbf P|\) is normalized by \(P_{\rm th,0}=\sum_s n_s(0)\sqrt{2m_sT_s(0)}\).

We find that while the nonlinear and linearized evolutions preserve the same collision invariants to roundoff, they follow different relaxation paths.
The nonlinear solution reaches \(A/A_0=0.3\) at \(t=7.65\), whereas the linearized solution reaches the same anisotropy at \(t=6.3\).
The difference is due to the transient collisional relaxation of a non-Maxwellian distribution, even with the same conservation of invariants.
Interspecies thermal equilibration is shown in \cref{fig:panel1} (d) compared at the same time $t=2.25/\nu^{ab}$ showing good agreement.
%

The perpendicular-plane diagnostic in Fig.~\ref{fig:panel1_3D} shows a second consequence of linearizing in a finite Cartesian basis.
It shows that a four-fold asymmetry develops when a linearized collision operator is used, which stems from the fact that truncated Hermite polynomial expansions are able to keep two-fold symmetries from symmetric distribution functions by initializing odd $n$ modes to zero as $H_n(-x)=(-1)^nH_n(x)$ but higher order asymmetries can persist, which get amplified when nonlinear contributions are neglected.
%
%
%
%
%
%
%
%
The relaxation and symmetry tests show that the use of invariant-conserving collision operators, as it is typically done in the simulation of weakly collisional far-from-equilibrium plasmas, is insufficient to reproduce the nonlinear relaxation path and rotational structure of the accurate Landau operator.

In this work, for the first time, we were able to implement and evolve the six-dimensional multispecies Landau integral by removing the $n^9$ storage barrier of using $n$ moments in velocity space, and the need to use toy models, therefore making large six-dimensional simulations feasible.
The model is derived from first principles using a cartesian Hermite formulation, yielding a single, simple to implement formula for the collision operator.
The relaxation tests show that, far from equilibrium, linearized operators can change the evolution of the distribution function and amplify angular projection errors.
This provides a practical route to nonlinear studies of fusion-edge transport, weakly collisional turbulence and non-Maxwellian space plasmas simulations.
%
%
%
%

Scripts and benchmark data are available at \href{https://github.com/rogeriojorge/nonlinear_collision_hermite}{github.com/rogeriojorge/nonlinear\_collision\_hermite}.
The code is written in Python using the JAX \cite{jax2018github} framework.

\begin{acknowledgements}
This work was supported by the National Science Foundation under Grant No. PHY-2409066. This research used resources of the National Energy Research Scientific Computing Center, a DOE Office of Science User Facility supported by the Office of Science of the U.S. Department of Energy under Contract No. DE-AC02-05CH11231 using NERSC award NERSC DDR-ERCAP0030134 and award FES-ERCAP0036936. This project received funding from the DOE Office of Fusion Energy Sciences under Award Numbers DE-SC0026040, DE-AC02-09CH11466, and DE-AC05-00OR22725. Support for this research was provided by the University of Wisconsin - Madison Office of the Vice Chancellor for Research with funding from the Wisconsin Alumni Research Foundation. The United States Government retains a non-exclusive, paid-up, irrevocable, worldwide license to publish or reproduce the published form of this manuscript, or allow others to do so, for United States Government purposes.
\end{acknowledgements}

\bibliography{references}



\end{document}